\documentclass[conference]{IEEEtran}
\IEEEoverridecommandlockouts
\usepackage{cite}
\usepackage{amsmath,amssymb,amsfonts}
\usepackage{algorithmic}
\usepackage{graphicx}
\usepackage{textcomp}
\usepackage{hyperref}
\usepackage{xcolor,soul}
\usepackage{caption}
\def\BibTeX{{\rm B\kern-.05em{\sc i\kern-.025em b}\kern-.08em
    T\kern-.1667em\lower.7ex\hbox{E}\kern-.125emX}}
\usepackage{enumitem}

\begin{document}

\title{Enhancing Cyber Threat Hunting: A Visual Approach with the Forensic Visualization Toolkit}

%A\author{\IEEEauthorblockN{ Jihane Najar}
%A\IEEEauthorblockA{\textit{dept. name of organization (of Aff.)} \\
%A\textit{name of organization (of Aff.)}\\
%ACity, Country \\
%Aemail address or ORCID}
%A{\href{https://orcid.org/0000-0002-3104-1169}{\protect\includegraphics[scale=0.5]{ORCIDiD_icon16x16.png}}}
%A\and
%A%A\IEEEauthorblockN{2\textsuperscript{nd} Given Name Surname}
%A\IEEEauthorblockA{\textit{dept. name of organization (of Aff.)} \\
%A\textit{name of organization (of Aff.)}\\
%ACity, Country \\
%Aemail address or ORCID}
%A\and
%A\IEEEauthorblockN{3\textsuperscript{rd} Given Name Surname}
%A\IEEEauthorblockA{\textit{dept. name of organization (of Aff.)} \\
%A\textit{name of organization (of Aff.)}\\
%ACity, Country \\
%Aemail address or ORCID}
%A
%A}

\author{
    \IEEEauthorblockN{
        Jihane Najar\IEEEauthorrefmark{1},
        Marinos Tsantekidis\IEEEauthorrefmark{1},
        Aris Sotiropoulos\IEEEauthorrefmark{1},
        and Vassilis Prevelakis\IEEEauthorrefmark{2}
    }
    \IEEEauthorblockA{
        AEGIS IT RESEARCH GmbH, Braunschweig, Germany\IEEEauthorrefmark{1}\\
        TU Braunschweig, Germany\IEEEauthorrefmark{2}\\
        Email: \{jnajar, m.tsantekidis, a.sotiropoulos\}@aegisresearch.eu\IEEEauthorrefmark{1}, prevelakis@ida.ing.tu-bs.de\IEEEauthorrefmark{2}
    }
}
\maketitle

\begin{abstract}
In today's dynamic cyber threat landscape, organizations must take proactive steps to bolster their cybersecurity defenses.
Cyber threat hunting is a proactive and iterative process aimed at identifying and mitigating advanced threats that may go undetected by traditional security measures.
Rather than waiting for automated security systems to flag potential threats, threat hunting involves actively searching for signs of malicious activity within an organization's network.
%Cyber Threat Intelligence (CTI), consisting of historical cyber threat data, holds a pivotal role in equipping security analysts with valuable insights into recent cyber threats and their mitigation strategies. 
In this paper, we present the Forensic Visualization Toolkit, a powerful tool designed for digital forensics investigations, analysis of digital evidence, and advanced visualizations to enhance cybersecurity situational awareness and risk management and empower security analysts with an intuitive and interactive tool.
Through practical, real-world scenarios, we demonstrate how FVT significantly amplifies the capabilities of cybersecurity professionals, enabling them to effectively identify, analyze, and respond to threats.
Furthermore, it is important to highlight that FVT has been integrated into, utilized, and continually enhanced within various EU-funded research projects over recent years.

\end{abstract}

%=============================================================================================================================
%=============================================================================================================================
\section{Introduction}
Cyber threat hunting has emerged as a vital discipline for detecting and countering advanced threats in today’s complex and constantly evolving digital environment. Organisations across the globe face the persistent challenge of safeguarding digital assets and sensitive information against an expanding spectrum of adversaries, ranging from highly sophisticated nation-state actors to opportunistic attackers. As both the scale and sophistication of threats grow, the demand for innovative methods of detection and response becomes increasingly urgent.

Visual analytics~\cite{andrienko2013visual,andrienko2018viewing} provides a framework in which computational methods reduce the need for direct human intervention, while still ensuring that visual representations are integrated throughout the analytical process~\cite{chae2017visual}. This integration is critical: in high-stakes scenarios, analysts must interpret complex information effectively to make timely and well-informed decisions. Visualisation thus plays a central role by presenting data in ways that support perception, comprehension, and cognitive reasoning.

In the context of cyber threat hunting, visualisation techniques are particularly valuable, as they enable the transformation of complex, high-volume data into actionable insights. By revealing hidden patterns, correlations, and anomalies, they act as a bridge between raw datasets and the analysts tasked with interpreting them.

This paper presents the Forensic Visualisation Toolkit (FVT), a novel solution that combines advanced visualisation techniques with data analytics to enhance the efficiency and precision of cyber threat hunting. The approach integrates diverse data sources and provides analysts with intuitive, interactive tools that support decision-making by offering a clear and comprehensive perspective on security incidents. Visual representation of data accelerates the detection of suspicious activity while deepening understanding of the broader threat landscape.

Specifically, the FVT brings together threat intelligence feeds, log data, and network traffic visualisations to deliver a unified view of an organisation’s security posture. Interactive dashboards and graphical interfaces allow analysts to explore relationships, correlate events, and detect indicators of compromise (IoC). This visual methodology not only expedites the detection of known threats but also facilitates the identification of novel attack patterns.

Crucially, the approach also addresses the human dimension of cybersecurity. By leveraging the natural visual processing abilities of analysts, the toolkit minimises cognitive load, enabling them to concentrate on critical threats and respond with agility.

Through real-world implementations, we demonstrate the practical value of visual cyber threat hunting. We discuss the associated challenges and opportunities, and provide guidance on how visual analytics can be embedded into existing security operations. Ultimately, we advocate for the broader adoption of visual analytics–driven cyber threat hunting as a means to strengthen organisational resilience in the face of an ever-changing threat landscape.

The rest of this paper is structured as follows: In Section ~\ref{rel_work}, we conduct a review of the current state-of-the-art concerning cyber threat hunting visualization techniques and tools.
Section ~\ref{design} introduces the design and architecture of FVT.
Section ~\ref{implementation} presents implementation details about FVT and real-world use-cases, where we apply FVT in a practical contexts.
Finally, in Section ~\ref{conclusion}, we bring our research to its conclusion.
%=============================================================================================================================
%=============================================================================================================================
\section{Related Work}
\label{rel_work}
This section presents an in-depth examination of relevant research and tools within the domain of cyber threat hunting focusing on visualization techniques.
Significant efforts have been invested into gaining a deeper understanding of this domain.
In~\cite{chae2017visual}, Chae et. al discuss the benefits and challenges of using visual analytics for intrusion detection, highlighting the need for integrating visualization tools into the cybersecurity workflow.
In~\cite{WenqiangVisualAnalytics}, a comprehensive survey of visual analytics is performed, which investigates how this approach is applied in various application domains.
Similarly, in~\cite{DamaševičiusVisualAnalytics}, the authors perform a wide-ranging overview of cyber security and digital forensics visualization works, offering a concise introduction to the subject and briefly mentioning some of the pertinent platforms.
Another approach in this direction is the work by Jang et al.\cite{jang2020fasttext}.
%In their research, they introduce a merged image-based malware classification framework, which encompasses local feature visualization, global image-based local feature visualization, and global and local image merging techniques.

Cyber threat hunting provides valuable insights into malware, attackers, and effective mitigation strategies.
However, several critical issues persist, including the handling of massive data volumes, the search for interesting patterns, and the automated integration of relevant information into security controls such as firewalls and intrusion detection systems, all of which require attention from the research community.
In addition to current research regarding cyber threat hunting, there is a number of tools, which provide the means to investigate and uncover signs of malicious activity (through memory or file analysis), while also allowing for the extraction of valuable digital artifacts.

Craig et al.~\cite{miles2014virusbattle} have introduced a web-based solution for conducting malware analysis.
This service enables the analysis of malware by examining its code and semantics.
Utilizing an interactive graph, it identifies connections between different malware strains, aiding in their classification.
However, this web service does not offer extensive analysis of cyber threat information.

STIXviz~\cite{STIXViz} is associated with the STIX project~\cite{STIX} and offers a graphical approach to represent STIX documents using a node-link graph.
It introduces a range of visualization options, including tree, graph, and timeline views, to facilitate in-depth analysis of STIX reports.
The tree view organizes STIX entities hierarchically, the graph view presents a force-directed graph representation of STIX documents, and the timeline view displays STIX data with timestamp information for chronological analysis.

Splunk~\cite{Splunk} is a commercial platform for processing unstructured log files into a centralized indexed database equipped with robust search, data processing, and visualization capabilities.
Splunk has gained popularity, particularly among system administrators and security analysts.
This arises from its ability to efficiently collect, index and aggregate data from diverse sources whether static or real-time, making it searchable from a centralized location.
Splunk serves as a data-centric solution that not only simplifies data exploration, transformation, and machine learning model training, but also offers an extensive selection of charting options, enabling users to interact with data both before and after the application of these models.

Plotly~\cite{Plotly} is another data visualization library and toolset that provides interactive customizable graphs, charts, and dashboards for data analysis and presentation.
It offers APIs and libraries for various programming languages like Python, R, and JavaScript.

Tableau~\cite{Tableau} stands as an interactive data visualization tool primarily tailored for Business Intelligence purposes.
Known for its range of visualization capabilities, Tableau facilitates the creation of customized visual representations and boasts compatibility with a diverse array of data formats and server connections. 
Offering an intuitive user interface and an extensive array of chart options, Tableau ensures a user-friendly experience.

%\begin{figure}[ht]
%\centering
 % \includegraphics[width=1\columnwidth]{Tableau Tool.png}
%\caption {Tableau User Interface}
%\label{fig:11}       % Give a unique label
%\end{figure} 

GRR Rapid Response~\cite{GRR} is an incident response framework developed by Google that specializes in remote live forensics. It relies on agents installed on target systems and a server to manage and interact with the agents. This setup enhances analysts' ability to quickly triage attacks and perform remote analyses.

Rekall~\cite{Rekall}, another Google framework, is an advanced forensics framework with incident response features, evolving from its initial focus on extracting digital artifacts from volatile memory to provide a comprehensive end-to-end solution for incident responders and forensic analysts. 

Another notable tool is Redline~\cite{Redline}, a free tool provided by FireEye~\cite{FireEye}.
Redline offers host-investigative capabilities to users in order to find signs of malicious activity through memory and file analysis and the development of a threat assessment profile.
With Redline, analysts can audit all running processes and drivers (from memory, file-system metadata, registry data, event logs, network information, services, tasks, and web history), and streamline memory and indicators of compromise analyses. 

Additionally, Volatility~\cite{Volatility} is an advanced memory forensics framework, which comes with a plug-in/add-on tool ecosystem, allowing it to offer additional features (e.g., malware analysis).
For instance, the VolatilityBot~\cite{VolatilityBot} offers automations pertaining to binary extraction and can help analysts in the first steps of performing a memory analysis investigation, while MalConfScan~\cite{MalConfScan} is a Volatility plugin that searches for malware in memory images and dumps configuration data.
%=============================================================================================================================
%=============================================================================================================================
\begin{figure*}[t!]
\centering
  \includegraphics[width=1\textwidth]{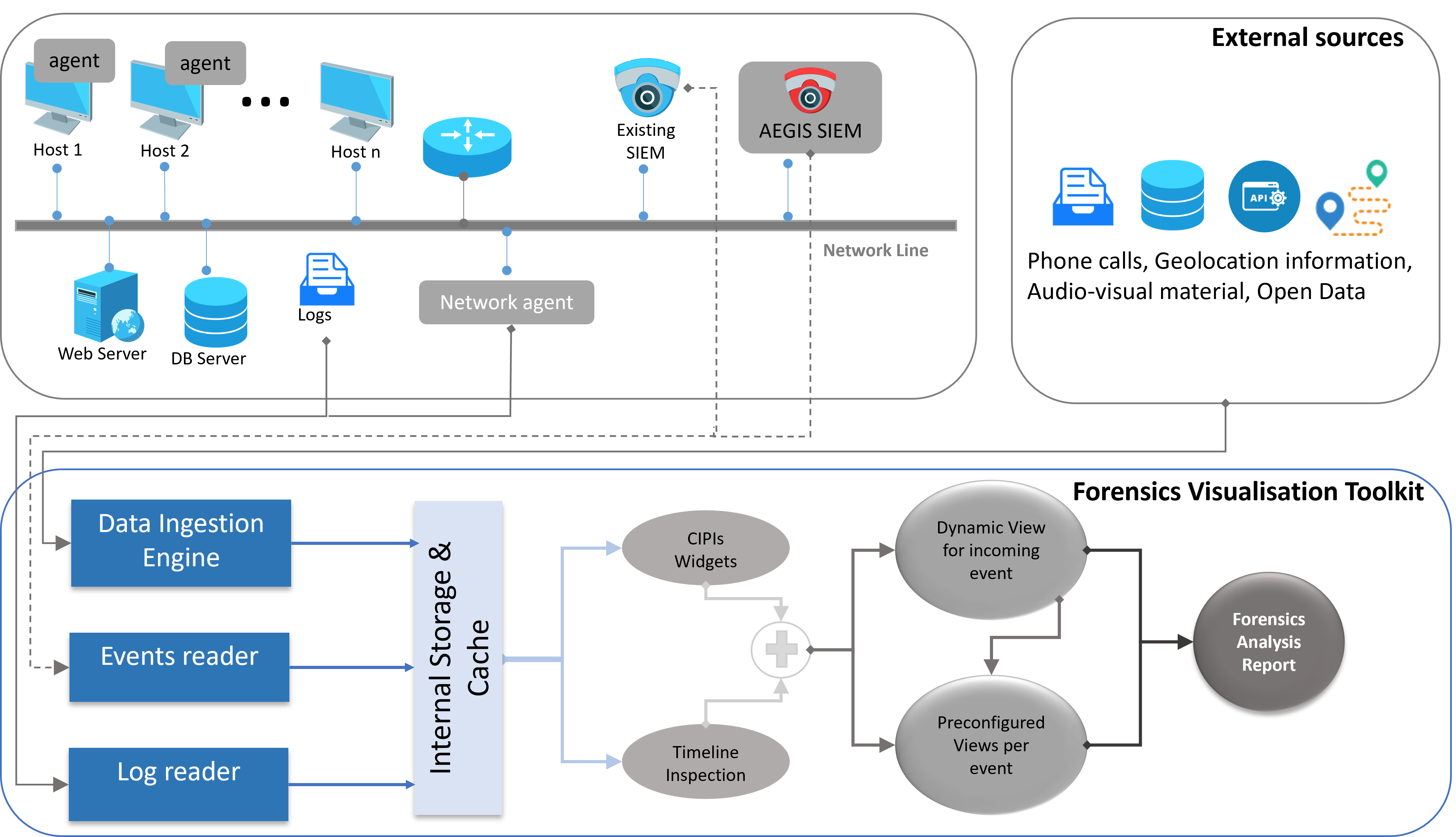}
\caption{FVT Internal Architecture}
\label{fig:FVT_Internal_Architecture}       % Give a unique label
\end{figure*}

\section{Design}
\label{design}
In the ongoing effort to strengthen defenses against continually evolving cyber threats, the array of cybersecurity tools continually expands.
However, the available tools for visualizing data have limitations.
These tools do not provide security analysts with the capacity to visualize extensive datasets from multiple perspectives and lack the flexibility required to implement robust defense strategies.
Due to the absence of a one-size-fits-all visualization tool resulting from the subjective nature of the human perception process, researchers either adapt existing visualization techniques to meet the unique demands of the cybersecurity domain, or they create innovative purpose-built approaches for visualizing cybersecurity data.

In this context, the FVT emerges as a fresh and tailored solution.
FVT is purposefully designed to empower digital forensics investigations, facilitate the in-depth analysis of digital evidence, and pioneer advanced visualizations that significantly enhance cybersecurity situational awareness and risk management.
In addressing the limitations associated with existing approaches, we introduce a novel approach to cybersecurity visualization, providing security analysts with a powerful toolkit to effectively navigate the complexities of the evolving cyber threat landscape.
FVT stands out in the field of cybersecurity by combining Host-based Intrusion Detection Systems (HIDS) with Network Intrusion Detection Systems (NIDS).
Furthermore, FVT serves as a network performance monitoring and diagnostic tool, providing a quick overview of an internal network's status and allowing operators to monitor network performance and flowing traffic. 
This fusion creates a robust cybersecurity tool with improved threat detection capabilities, enhanced security visibility, and exceptional situational awareness.

What distinguishes FVT are its three fundamental pillars:

\begin{itemize}
    \item \textbf{Intuitive and Detailed Visualizations}: FVT redefines the landscape of cyber and digital forensics analysis by offering intuitive and detailed visualizations in real-time. Unlike traditional visualization tools, which tend to be vertically integrated, FVT allows for the simultaneous use of multiple real-time views of the same data. %This innovation empowers analysts to gain multifaceted perspectives and insights into cybersecurity threats, providing a more holistic understanding of the threat landscape.
    
    \item \textbf {Customized Implementation for Critical Infrastructure (CI) Forensics Analysis: }: Recognizing the unique challenges posed by critical infrastructure environments, FVT is designed for adaptability. It empowers organizations to tailor forensics services to their specific CI, gathering effective data for post-incident security analysis, seamlessly integrating post-incident analysis into a defense-in-depth strategy, enabling real-time forensics analysis, correlating forensic data from diverse sources, and accommodating various views. %While cyber forensics is a mature domain, each organization must align forensics services with its CI to effectively address environment-specific challenges. FVT conducts an in-depth analysis of the specific CI and defines a set of Critical Infrastructure Performance Indicators (CIPIs) essential for sufficient deployment monitoring.
    
    \item \textbf{Innovative Forensics Services}: In addition to its visualization capabilities, FVT introduces innovative forensics services. These services encompass timeline analysis, preconfigured views, and best practices. The timeline analysis capability allows for the comprehensive examination of heterogeneous events through advanced visualizations, facilitating a deep dive into the sequence of events leading up to a security incident. Preconfigured views automatically adapt visualizations based on past situations, streamlining the analysis process.
\end{itemize}

%\begin{figure}[ht]
%\centering
%  \includegraphics[width=0.7\columnwidth]{FVT Technology Stack.png}
%\caption{FVT Technology Stack}
%\end{figure}

%Its innovations include: (i) the incorporation of both physical and cyber forensics services and algorithms; (ii) timeline analysis of a large number of heterogeneous events via advanced visualizations; (iii) preconfigured views that provide automatically adapted visualizations based on similar past situations; and (iv) threat hunting capabilities empowered by correlation algorithms, enabling almost real-time mitigation of security incidents.

By offering interconnected, interactive visualizations of security-related events and performance indicators, FVT enhances end-users' situational awareness and aids in identifying patterns or correlations that foster the threat-hunting process.
Operators can proactively identify abnormal situations, take appropriate actions to prevent threats, or trace the root cause of security incidents through after-the-fact analysis.
FVT offers operators:

\begin{itemize}
    \item \textbf{Time Travel Capability:} The ability to 'travel back in time' and compare the current situation with past events, enabling the identification of threat patterns and informed response strategies based on historical outcomes.
    
    \item \textbf{Customizable Views:} The ability to adapt the display of information based on previously encountered situations, streamlining analysis and reusing effective visualization configurations for new cases.
\end{itemize}

Combined, these two aspects of FVT allow analysts to quickly gain a solid understanding of an event and benefit from existing knowledge stored in the repository. Information is automatically presented to enhance situational awareness, reducing the need for technical support at the end-user level, which can be particularly beneficial in forward deployments.

\subsection*{Architecture}
%The Forensic Visualization Toolkit (FVT) is a pivotal component of our research, designed to seamlessly integrate with the world of Cyber Threat Intelligence (CTI) and enhance the capabilities of cybersecurity professionals.\\
%At its core, FVT's frontend leverages the Angular.io framework, complemented by a versatile set of visualization libraries. These libraries are selected and interchangeably used to address the specific needs and particularities of the data under analysis (Figure \ref{fig:2}). They include a wide array of visualization techniques, including interactive timelines for time-dependent data, graph-based visualizations, chord diagrams, geospatial representations, and more.

%AIn terms of data management/data storage capabilities, FVT can operate with its own storage facility mainly driven by the ElasticSearch search engine. Additionally, it integrates with various other data sources, whether they are data streams such as Kafka, or batch data accessed via RESTful APIs, external ElasticSearch, MongoDB, and more. This flexibility is possible due to a number of tailor-made converters, which can request the data from the respective data sources and transform it to a suitable format for FVT’s visualisation widgets. These transformations, along with internal configuration options and business logic activities, are managed within FVT's middleware application, which is developed using Node.js.

The frontend of FVT is based on Angular.io framework in conjunction with a set of visualisation libraries that are interchangeably used to cover specific needs and particularities of the analysed data.
These include interactive timeline representations for time-dependent data, graph-based visualisations, chord diagrams, geospatial representations, etc.

In terms of data storage capabilities, FVT can operate with its own storage facility [(mainly Elasticsearch~\cite{gormley2015elasticsearch} (ES) engine] and/or also use multiple other sources of data, be it data streams (e.g. Kafka~\cite{garg2013apache}) or batch data (e.g. via RESTful APIs, external ES, MongoDB, etc.).
This flexibility is possible due to a number of tailor-made converters, which can request the data from the respective data sources and transform it to a suitable format for FVT’s visualization widgets.
These transformations together with internal configuration options and business logic activities take place in FVT’s middleware application built on Node.js.

FVT offers two distinct modes of deployment.
It can be installed as a standalone application with moderate hardware and software requirements.
Alternatively, it can be deployed as a containerized application using Docker, providing a highly configurable and easily integrated solution for existing environments.

Figure~\ref{fig:FVT_Internal_Architecture} depicts the internal architecture of FVT and how it can ingest data from a local network and other external sources and visualize them.
FVT serves as the central hub for collecting data from monitored components and assets, including computers, network devices, switches, and more.
This data enables cybersecurity professionals to thoroughly investigate and detect potential abnormal system operations, all within the purview of CTI, thus strengthening their cyber defense strategies.

%=============================================================================================================================
%=============================================================================================================================
\section{Implementation}
\label{implementation}
FVT can inform end users on security/privacy levels of the operating infrastructure while providing warnings and assisting them in handling security and privacy related incidents.
Such an advanced environment must be presented in an easy to use and understandable way to the end users so as to help them embrace the offered features and understand the benefits of using such an ecosystem by removing the complexity of the underlying technologies.
Therefore, the interface elements follow an intuitive and flexible design that can be adjusted to any device.
This approach allows the end users to manage the monitoring of their industrial infrastructure, inspect the incidents that might occur and view generated statistics and insights.
The securely stored data, analytics results, and user transactions are presented by advanced visualisations that can foster a real-time threat hunting processes (empowered by correlation algorithms for almost real-time mitigation of security incidents) and post-mortem digital forensics investigations (including physical security-like incidents).

\begin{figure*}[t!]
\centering
  \includegraphics[width=1\textwidth]{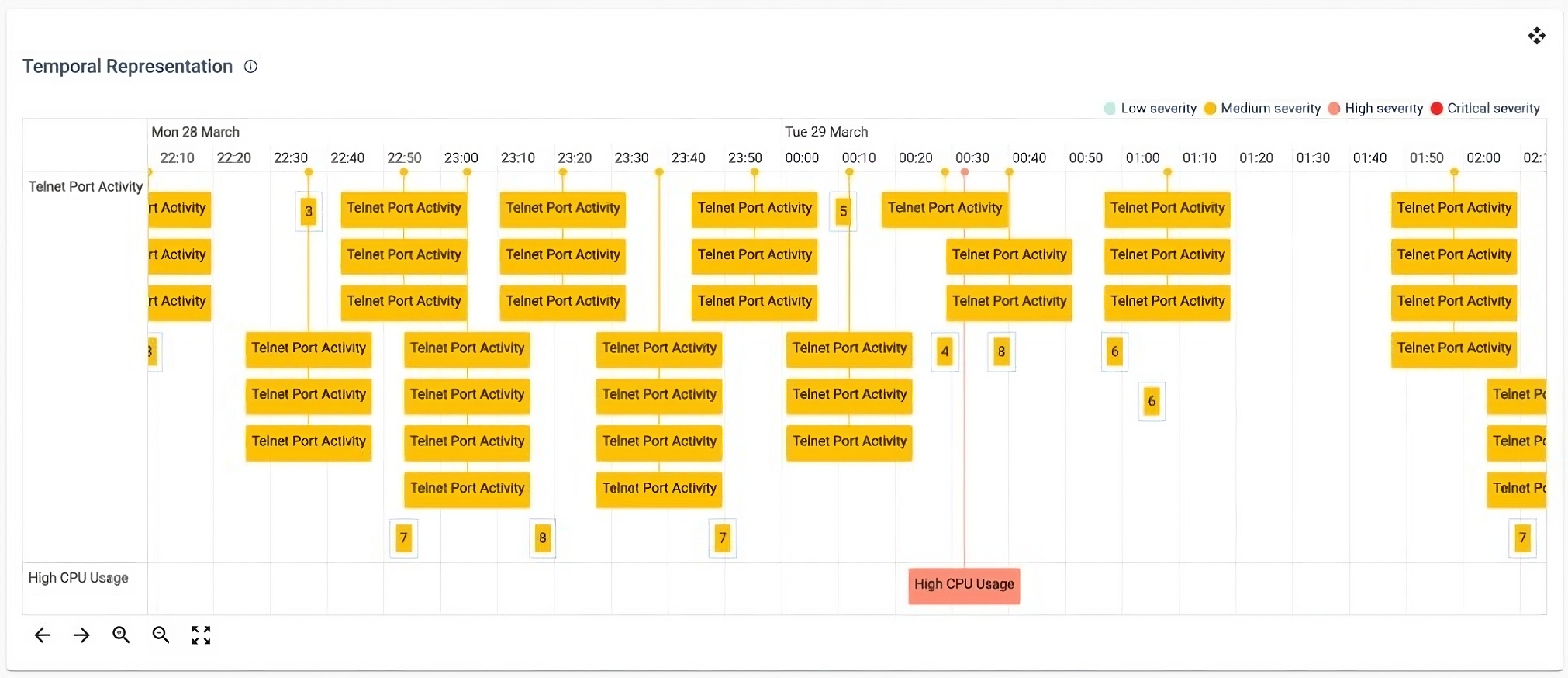}
\caption{FVT Timeline Analysis}
\label{fig:temporal_representation}       % Give a unique label
\end{figure*}

FVT can also be adjusted to act as a Network Performance Monitoring and Diagnostic (NPMD) tool that is deployed close to an unmanaged machine network (e.g., a production line of a factory).
In this case, it provides historical, real-time and predictive views into the availability and performance of the network and the application traffic running on it, by leveraging a combination of packet data, flow data and infrastructure metrics.
Such functionality also provides the diagnostic workflows and forensic data to identify the root causes of performance degradations, increasingly through artificial intelligence for IT operations (AIOps) functionality.

The internal engine of FVT can ingest several types of data from a number of heterogeneous sources and components:
\begin{itemize}
    \item \textbf{Logs:} There is a monitoring agent deployed in each monitored host and/or more centrally in a network that produce a number of text-based logs.
    
    \item \textbf{Security Information \& Event Management (SIEM) systems:} They aggregate data from various sources, analyse real-time and historical security-pertinent events, and generate alerts based on predefined rules to detect and respond to security threats.
    
    \item \textbf{Cyber Threat Intelligence (CTI) platforms:} They aggregate, analyse and disseminate CTI indicators about potential and current cyber threats, providing enhanced situational awareness, early warning on attacks, and an up-to-date view of the evolving threat landscape.
    
    \item \textbf{Endpoint Detection \& Response (EDR) tools:} They assist in investigating and detecting suspicious activities (and traces of such activities) and other problems on individual hosts/endpoints within a network (e.g., antivirus, rootkit scanners,  HIDSs, etc.) and enable rapid response to potential security incidents.
    
    \item \textbf{Network Detection \& Response (NDR) tools:} They are centered on scrutinizing network traffic for signs of malicious actors and unusual activities, while facilitating the ability to respond swiftly to identified cyber threats within the network.
    
    \item \textbf{Deception tools:} They emulate actual production systems to mislead and detect malicious actors within a network. These tools (e.g., honeypots) create deceptive elements, such as decoy systems, data, or credentials, to divert and lure attackers away from actual critical assets. They provide insights of persistent threats, previously undiscovered attacks, vulnerabilities, and malware.
\end{itemize}

In order to deal with the multitude of components from where FVT can integrate data, it supports a number of possibilities, depending on the infrastructure and the configuration of the component:
\begin{enumerate}[label=\roman*)]
    \item Receive data from each component individually (via RESTful APIs, internal databases. etc.) and store it in FVT's own internal ES instance. FVT probes each component separately to get any updated data. This option is more suitable for historical views rather than real-time monitoring.
    
    \item Receive data from an external central database, where all components store their data, following their separate ways to do it.
    
    \item Receive data from each component individually through a message broker/aggregator and then store it in FVT's own internal ES instance. Each component (including FVT) communicates with a central point leveraging the publish/subscribe model (e.g., RabbitMQ, Kafka, etc.), using an appropriate messaging protocol of the tool owner's choice. In this way, all data is consumed automatically as soon as it is made available. Other than historical views, this option is also suitable for (near) real-time monitoring.
\end{enumerate}

\begin{figure}[ht]
\centering
    \includegraphics[width=1\columnwidth]{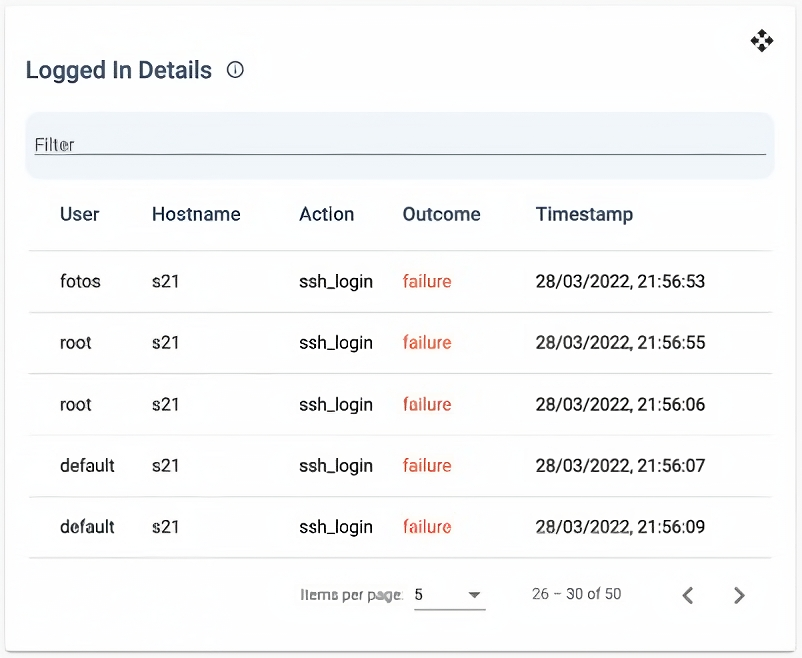}
\caption{Logged-in users}
\label{fig:users}       % Give a unique label
\end{figure} 

\begin{figure}[t!]
\centering
    \includegraphics[width=1\columnwidth]{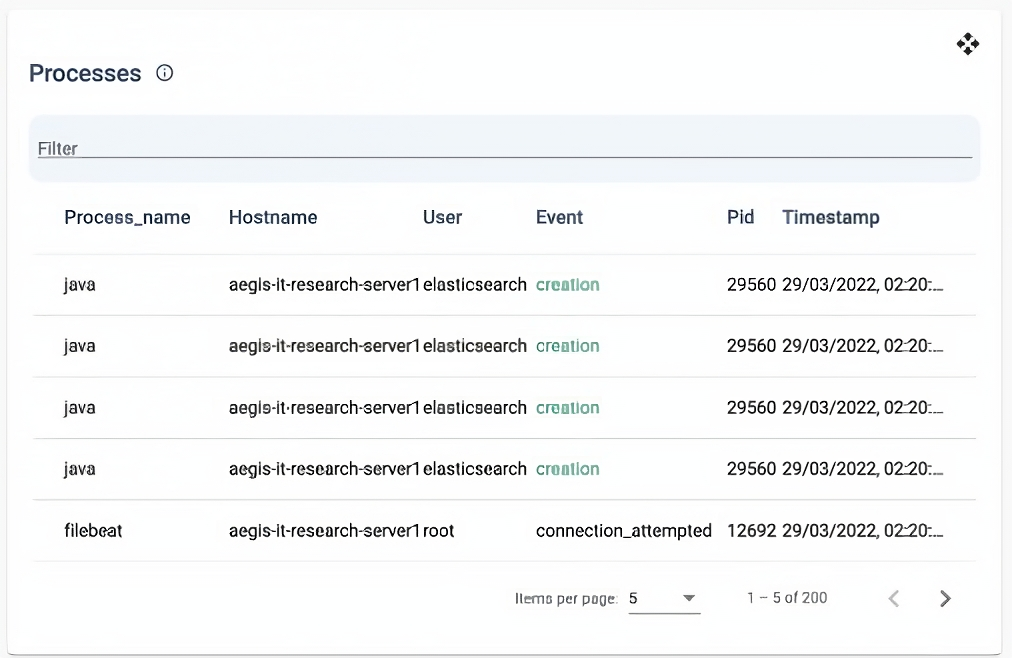}
\caption{Processes}
\label{fig:processes}       % Give a unique label
\end{figure}

\begin{figure}[t!]
\centering
    \includegraphics[width=1\columnwidth]{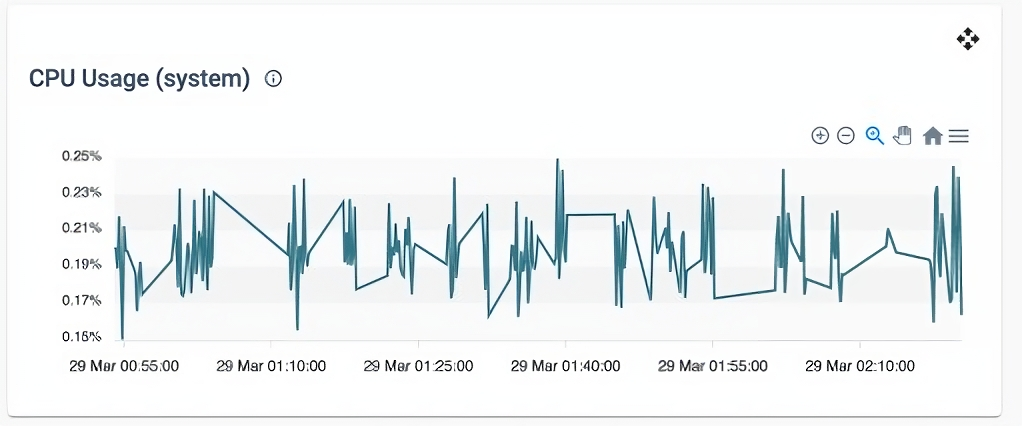}
\caption{CPU usage}
\label{fig:CPU_usage}       % Give a unique label
\end{figure}

\begin{figure*}[t!]
\centering
  \includegraphics[width=1.5\columnwidth]{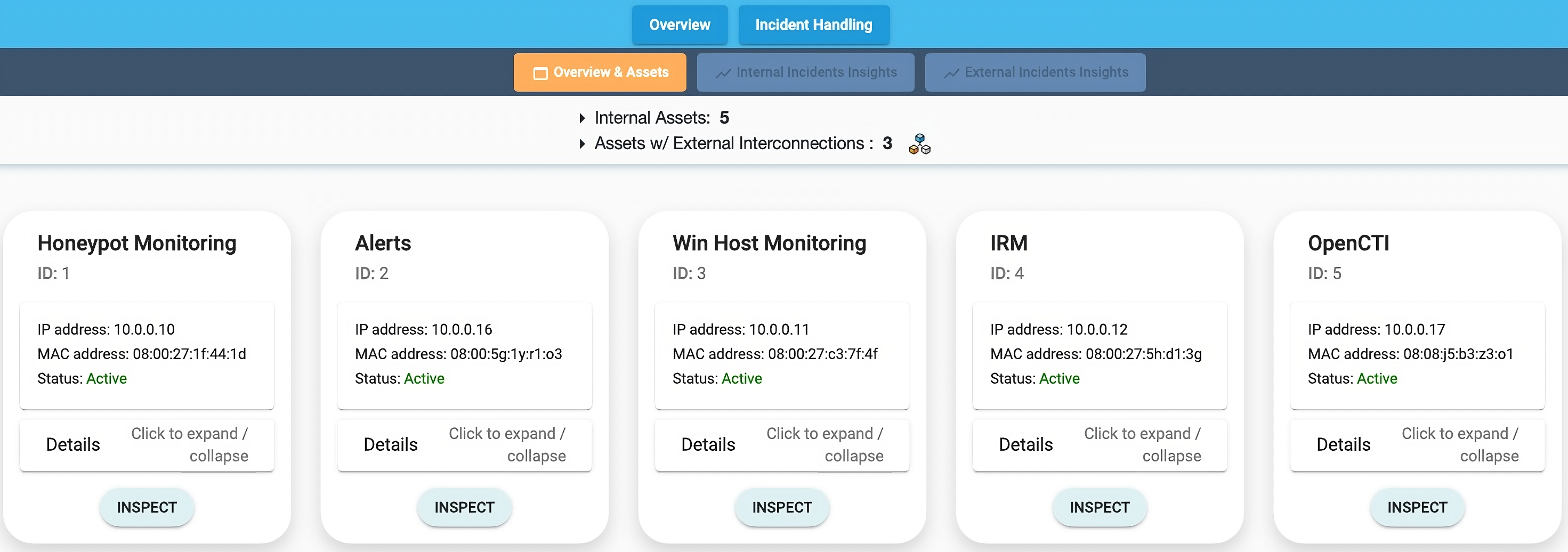}
\caption{FVT Tiles}
\label{fig:FVT_tiles}       % Give a unique label
\end{figure*}

In any case, since the data is usually in different formats, FVT uses custom-made converters in order transform it to a suitable format for the visualization widgets.  
The user can then see a multitude of details such as:
$(a)$ a textual list of alerts that the host has produced, which is also depicted in a temporal graphical representation for a selectable time period at the Timeline widget (Figure~\ref{fig:temporal_representation}), $(b)$ the users that have logged in and their sessions (Figure~\ref{fig:users}), $(c)$ the processes that are executed in the host machine (Figure~\ref{fig:processes}), and
$(d)$ its CPU usage (Figure~\ref{fig:CPU_usage}).
All text-based data can be visualised in the Timeline widget similar to Figure~\ref{fig:temporal_representation}.
A number monitored assets can be displayed as well in distinct tiles (e.g. Honeypot, CTI platform, Host monitoring, etc.) as shown in Figure~\ref{fig:FVT_tiles}.
\begin{figure}[t!]
\centering
  \includegraphics[width=0.9\columnwidth]{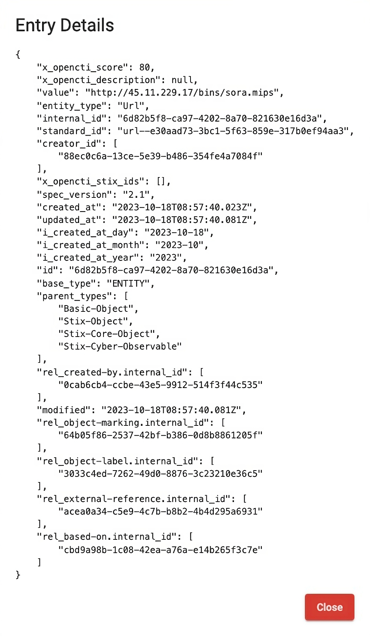}
\caption{Example of a cyber observable}
\label{fig:OpenCTIView}       % Give a unique label
\end{figure}

As mentioned in Section~\ref{design}, all these tiles can be combined in a preconfigured view or can be moved around the dashboard, in order to fit the needs of the user.
The user interface can also be customised so it is presented in different colors.

%\begin{figure}[ht]
%\centering
 % \includegraphics[width=1\columnwidth]{alerts.png}
%\caption{List of alerts}
%\label{fig:alerts}       % Give a unique label
%\end{figure}

\subsection*{Real-life Deployment }

FVT is a commercial tool, however it has been introduced in, used in and upgraded within several EU-funded research projects throughout recent years.
In these projects, a combination of the capabilities depicted in the Figures above and below has been implemented.

\begin{figure*}[t!]
\centering
  \includegraphics[width=1.3\columnwidth]{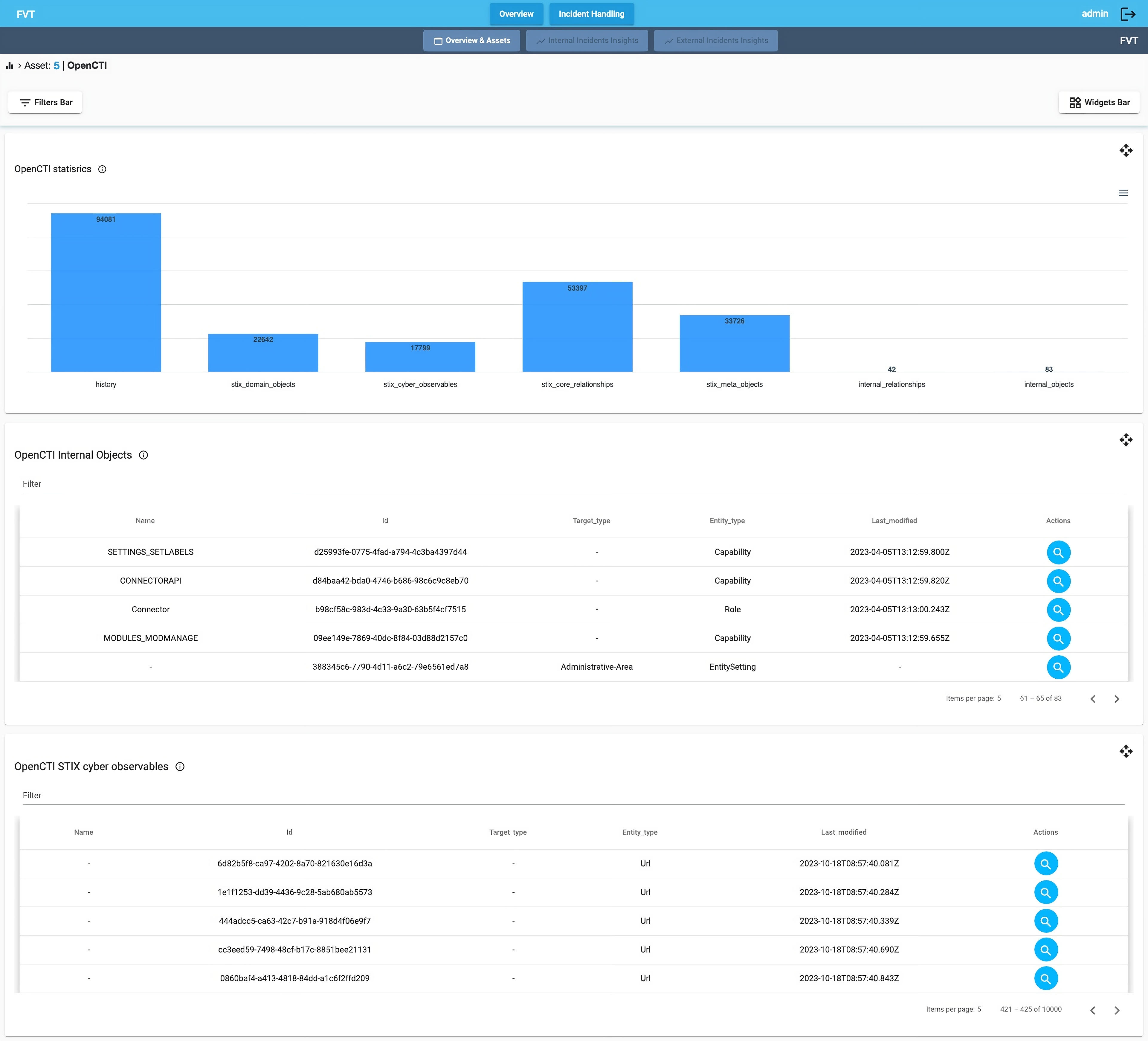}
\caption{OpenCTI FVT dashboard}
\label{fig:openCTI_main_DASHBOARD}       % Give a unique label
\end{figure*}

In the context of the JCOP project, the FVT dashboard has been configured to illustrate information ingested from the OpenCTI platform (Figures~\ref{fig:FVT_tiles} and~\ref{fig:openCTI_main_DASHBOARD}) among others.
There is a barchart widget displaying all available indices/topics within the OpenCTI platform. This visual representation allows users to grasp the volume of entries across various categories, such as cyber observables, domain objects, relations, etc. 
Below this widget, as presented in  figure ~\ref{fig:openCTI_main_DASHBOARD}, there are two tables: one dedicated to the internal objects within the cyber threat intelligence ecosystem and another for cyber observables. These tables, equipped with preconfigured views offered by the FVT, allow the user to customize and view only the topics of their interest.
The tables feature server-side pagination, providing a robust interface for users to navigate, search, and analyze the indicators present in the OpenCTI indicators.
In the Internal Object's table, the user can investigate data points that indicate the presence of a cyber threat. These could range from specific indicators like IP addresses, domain names, hashes, etc., to relationships between entities within the OpenCTI ecosystem. Attributes, which represent characteristics or properties associated with different objects, as well as entire STIX-formatted objects that represent entities related to specific cyber threats, are also accessible.
In the second table, dedicated to cyber observables, users can gain a high-level view of STIX-formatted data instances or patterns relevant to cyber threats. This includes file hashes, IP addresses, domain names, network traffic patterns, email addresses, and other observable data that can be used for identifying and characterizing cyber threats.
Figure ~\ref{fig:OpenCTIView} presents an example of a cyber observable presented in STIX format. Accessible by clicking the magnifier icon within the Actions column of the table, this modal provides an overview of the specific cyber observable.
If the user requires more information about a specific IoC, they can be redirected to OpenCTI platform where the can find more details about it.

%\begin{figure}[ht]
%\centering
 % \includegraphics[width=1\columnwidth]{available_hosts.jpg}
%\caption{List of available hosts}
%\label{fig:available_hosts}       % Give a unique label
%\end{figure}
%Figure~\ref{fig:available_hosts} shows how the user can select from a number of available monitored hosts.

%Additionally, the FVT stands as the main dashboard of the platform of the projects in several cases, collecting input from a number of backend services and modules and displaying it in an intuitive manner.

In the context of the HEIR project, FVT has been adapted to provide intuitive visualisations and forensic analysis through the User Interface of the developed platform.
HEIR aims at boosting the overall level of digital health security in Europe by developing a novel platform that provides a thorough threat identification and cybersecurity knowledge-base system that focuses on depicting the landscape of cyberthreats for ICT-based healthcare ecosystems, detailed cybersecurity assurance statuses, and their evolution over time.
In addition, HEIR intents to establish good security practice in regulatory frameworks across Europe to reduce market access limitations, conflicting requirements and unnecessary administrative burdens. 

In this context, FVT aims to assist healthcare infrastructures when identifying possible threats. This is a challenging task for several reasons, such as: 
\begin{itemize}
    \item the sophisticated technologies often used when planning and executing cyber attacks,
    
    \item the growing variety of heterogeneous data that need to be collected, filtered, analysed and secured,
    
    \item the velocity of this data that tends to be high; healthcare infrastructures involve thousands of patients and each one of them may use several services,
    such as diagnostic exams or medical follow-up during admissions, resulting in activity that needs monitoring,
    
    \item the size of data that need to be processed grows further as healthcare infrastructures consist of several departments, resulting in thousands of medical devices and computer systems,
    
    \item the large number of devices and security data that need to be monitored. 
\end{itemize}

%\begin{figure}[ht]
%\centering
 % \includegraphics[width=1\columnwidth]{processes.png}
%\caption{Processes}
%\label{fig:processes}       % Give a unique label
%\end{figure}

FVT offers a solution that can handle large number of heterogeneous events and provide intuitive visualisations for IT and non-IT experts that reveal hidden relationships and insights.
As a result, FVT can increase the efficiency and effectiveness of Critical Infrastructure domains, such as healthcare, in several phases, namely:
\begin{itemize}
    \item offering real-time and post-mortem forensic analysis,
    
    \item processing and combining data of various modalities (e.g., connected departments per hospital, attacks number, etc.) by automating the most time-consuming tasks,
    
    \item automatically producing information on the hospitals (e.g., critical events identified) and their interactions, etc,
    
    \item intuitive analysis, assessment and finetuning of FVT outputs by allowing users to provide expert knowledge (if, and when, needed).
\end{itemize}

\section{Conclusion}
\label{conclusion}
In this paper, we present the FVT as a multifaceted solution, combining data analytics, advanced visualizations, and real-time threat mitigation.
Its adaptability to various scenarios, from network monitoring to forensic data analysis, showcases its versatility in the ever-evolving field of cybersecurity.
Its real-life implementation exemplifies its capacity to empower end-users by providing an easy-to-use, intuitive interface.
This interface enhances their ability to monitor their industrial infrastructure, investigate incidents, and gain valuable insights through advanced visualizations.
FVT equips organizations with the tools and insights they need to stay ahead of cyber threats and safeguard their digital assets.

FVT is not intended to substitute the components from which it integrates all data.
Its role is to be an umbrella, under which all these components can present a initial set of information to the user, in an intuitive manner.
The user has all this information aggregated in one dashboard, with a common User Interface, enabling them to swiftly take informed decisions and actions, based on events and alerts produced by the monitored infrastructure.
In case they require additional details about an event/alert, they can always be redirected to the specific component that produced and disseminated the relevant data.

As future work, our primary focus will revolve around conducting a comprehensive comparative analysis to assess the performance of FVT in contrast to similar tools. This examination will encompass multiple dimensions, emphasizing not only the varying types of data involved—ranging
from diverse digital evidence encountered in file systems to network logs—but also exploring data from different devices and digital environments. Specifically addressing intuitiveness, the experiment will integrate user experience testing methodologies. This will involve cybersecurity professionals
and digital forensics investigators engaging in simulated scenarios to evaluate FVT’s ease of use, interface navigation, and overall user experience. Additionally, we’ll incorporate task-based evaluations and user feedback to iteratively enhance FVT’s intuitiveness. The scale of this experiment will vary, encompassing small-scale simulations to extensive analyses of larger and more complex datasets. This holistic approach aims to not only ascertain FVT’s strengths and capabilities but
also guide future developments and optimizations, ensuring it remains an intuitive and efficient tool for cybersecurity professionals.

\section*{Acknowledgements} This work is supported by the following European Union-funded projects: a) AI4HEALTHSEC (Agreement No.: 883273), b) PHOENI2X (Agreement No.: 101070586), c) JCOP (Agreement No.: INEA/CEF/ICT/A2020/2373266) and d) CyberSecPro (Agreement No.: 101083594)

\bibliographystyle{spmpsci}  
\bibliography{conference_101719}  

\end{document}